# Link Analysis for Communities Detection on Facebook

**Mohamed Adnane Mellah[1], Abdelmalek Amine[2*], Reda Mohamed Hamou[3], A.V. Senthil Kumar[4]**

[1,2,3]GeCoDe Laboratory, Department of Computer Science, Tahar Moulay University of Saida, Saida, Algeria.

[4]Director, Department of MCA, Hindustan College of Arts and Science, Bharathiar University, Coimbatore–28, Tamil Nadu, India

Email id : *amine_abd1@yahoo.Fr; [1]mohamed.mellah@hotmail.fr; [3]hamoureda@yahoo.fr; [4]avsenthilkumar@yahoo.com

**Abstract**

Social networks have become a part in the daily life of millions of users, which offer wide range of interests and practices. The main characteristic of social networks is its ability to gather different individuals around a common point of view or collective beliefs. Among the current social networking sites, Facebook is the most popular, which has the highest number of users. However, in Facebook, the existence of communities (groups)is a critical question; thus, many researchers focus on potential communities by using techniques like data mining and web mining. In this work, we present four approaches based on link analysis techniques to detect prospective groups and their members.

## 1. INTRODUCTION

Facebook and Twitter are social actors similar to individuals or organizations who are linked together by social interactions. It describes a dynamic social structure by a set of nodes and links. The analysis of socialnetworks, mainly based on graph theories and sociological analysis, aims to study different aspects of these networks. The main aspects are community detection, identification of influential actors, and the study and prediction of the evolution of networks.

The discovery of communities [1, 2, 3, 4, 5,6] is an important problem in social network analysis, where the goal is to identify the groups (communities) as well as their members and the ones that belong to several communities. Researchers focus on different methods to detect communities from social networks; the majority of methods supposes that communities are separated where each member of these communities constitutes a node that is categorized under one label. In the real world, a member can be interested by various topics—for instance students can belong to more than one community. Assign multiple labels to the same node is the best representation of the properties of a social network. The most common definition of a community is as follows: ‘A community is a part of a graph where the nodes are strongly related together compared to the other nodes of the same graph'; numerous approaches to detect communities in a social network were proposed in the past.

In this work, we present four approaches which are based on link analysis algorithms for the detection of communitieson Facebook, which are PAGERANK, HITS, PHITS and SALSA. Practically, the social communities are manually managed by their administrators, and any user can join these communities creating links to these pages. If the ranking attributed by these algorithms to a node is higher, it is probable that the node constitutes a page which determines a community. In the second step, we verify if this node does not contain outgoing links; in this case, it constitutes a page of group and the ingoing links towards this page are the members of this community which share common characteristics and interests. It is to be noted that Facebook does not permit the administrators of communitiesto send invitations and create outgoing links from the pages of the communities, and for this reason, we verified the presence of outgoing links of these nodes.

The rest of this paper is organized as follows: Section 2 presents related works. Section 3 deals with algorithms and theory of link analysis. Section 4 shows the methodology of research. Section 5 shows an experimental analysis. Finally, Section 6 concludes this paper.

## 2. Related Works

In the domain of recognition of web communities, many studies are available. In Gibson et al [7], the hyperlink is used as a basis for reasoning. The major contribution in this area is the HITS algorithm of Kleinberg [7] which defines the notions of authority and hubs, structuring a community around a given topic. Imafuji and Kitsuregawa [8] suggest that a page belonging to a community (if this page is referenced primarily from interior of the community more than for its exterior) uses the maximum number of algorithms to isolate the nodes belonging to the same community. Based on the proposed algorithm by Flake et al. [9], Dourisboure et al. [10] identified a graph of Web communities as heavy subgraphs with a bipartite in this graph. The bipartite graph represents,on the one hand, the interests of the community (authorities according HITS) and, on the other hand, those who cite the community (hubs). This method allows highlighting the potential of sharing similar interests by several communities of actors or vice versa.

These approaches provide an advanced analysis of the links between different pages structuring a thematic community, but do not bring users together by their interests or activities: sharing hyperlink is no longer necessarily the basis of community activity in social exchanges collaborative Web (content evaluation by the user, affixing tags, etc.).

## 3. Link Analysis Algorithms

From its origins in the bibliometric analysis, the analysis of reasons for referrals (link analysis) has come to play an important role in the salvaging of modern information. The algorithms of link analysis [11, 12, 13, 15, 16, 17] have been successfully applied to Web hyperlinking data to identify sources of official information and citation data for the most important items. Currently, with the conventional classification techniques, link analysis is based only on some research engines on the Internet. An important feature of the World Wide Web is its dynamic nature; the references can be modified so that it becomes inaccessible or simply not found by the search engine. If link analysis provides a notion of robustness in such a context, it is natural to ask whether robustness means being stable to perturbations of the link structure. Indeed, a completely unstable search engine that changes its results every day can cause lots of confusion; it is for this reason that several algorithms and strategies have emerged.

### 3.1 Preliminaries

Let us consider a set of Web documents interviewing each other; this collection can be seen as a directed graph. Links analysis algorithms construct the adjacency matrix that represents the reflection of the graph based on the model of citation that is used. For example, the bond between two documents i and j is represented by the value 1 of the element $W_{ij}$. The most interesting pages can be then extracted by computing the eigenvectors of the system. Based on the meaning of Kleinberg, these pages can be divided into two categories:

**Hubs**: pages containing little relevant information, but many hyperlinks.

**Authorities**: pages with few links, but a lot of relevant information.

### 3.2 The INDEGREE algorithm

The main idea of this algorithm[19] is very simple; such pages are considered as most popular if they have much more incoming links than other pages. Studies have shown that the INDEGREE algorithm is not sophisticated enough to capture the authority of a node. In graph G: for each node i, $A_i = |B(i)|$.

### 3.3 The PageRank algorithm

The PageRank algorithm proposed by L. Page and S. Brin[13] allows the assignment of a reputation score for each page found on the Internet. This algorithm quantifies the reputation of a page by counting the number of hyperlinks that point to it: thus a page with many incoming links is considered very popular and therefore enjoys a high reputation. However, a hyperlink from page i to page j is considered as a vote of i for the page j: for each Web page i referenced by Google, a local vector reputation score $c_i$ is calculated, where ci, j = 0 if there is no link from i to j and ci, j = 1/Li if at least one link exists (Li is the number of links on the page i). The PAGERANK Ri of a page i is calculated by the sum of PageRank (weighted by the inverse of the number of links) pages which point towards B.

$$Ri = \sum_{j \in B} \frac{Rj}{|Lj|} \qquad (1)$$

The formula for calculating the PAGERANK of a page is recursive; the PAGERANK is approximate within an iterative method. We initialize the algorithm by a non-zero constant value of PAGERANK for each page; and at each iteration we re-compute the PAGERANK of each page using the formula. Iterations are repeated to achieve a convergence of PAGERANK values.

---

PAGERANK

---

1 Initialize all Ri to 1
2 Repeat until the Ri converge:
3 if there are other nodes to brows

$$Ri = \sum_{j \in B} \frac{Rj}{|Lj|}$$

4 *PAGERANK ()*, calling the recursive function

---

### 3.4 The HITS algorithm

An algorithm called HITS proposed by Kleinberg [15] will be able to identify the best hubs and authorities in a hyperlinked collection. This algorithm exploits the structure of the web graph. Each document is seen as a directed graph node, and any link between two documents is interpreted as an edge between the two nodes. Based on a specific query, called σ, the algorithm first creates a subgraph. It then calculates the weight of the hubs and authorities for each node Sσ. The principle used by the HITS algorithm is as follows: a document has a high weight authority if is pointed to many documents with high hub weight and vice versa, and a document has a high hub weight if it points to many documents with high authority weight. More specifically, starting from a hyperlinked set of documents, the HITS algorithm builds the directed graph associated with the collection. Ideally, the collection S must satisfy the following properties:

(i) S is relatively small.

(ii) S contains many relevant pages.

(iii) S contains most of the best authorities.

The graph is represented by an adjacency matrix W n × n, where n represents the number of used materials. The element Wij takes the value 1 if there is an edge between nodes i and j in the directed graph, and 0 otherwise. Generally, the third condition is not satisfied and the S collection should be extended by exploring a number of links of the graph (Kleinberg, 1998). The algorithm can then calculate the relationships of mutual reinforcement between hubs and authorities iterating rules by following the update:

$$h_i^{(k+1)} = \sum_{j:(j \to i)}^{n} a_j^{(k)} \; ; \; a_i^{(k+1)} = \sum_{j:(i \to j)}^{n} h_j^{(k)}$$

Where « $i \to j$ » means that the documenti points to the documentj.

HITS

1 Initialize all weights to 1
2 Repeat until the weights converge:
3 for every hub $i \in H$
$h_i = \sum_{j \in F(i)} a_j$
4 for every *authority* $i \in A$
$a_i = \sum_{j \in B(i)} h_j$
5 Normalize

### 3.5 The SALSA algorithm

‘The Stochastic Approach for Link Structure Analysis’ (SALSA)is an algorithm based on the theory of Markov chainsproposed by Lempel and Morgan [16]. The algorithm uses the properties of a random walk performed on a collection of hyperlinked documents. Similar to the Kleinberg algorithm, SALSA starts by building a basic collection ‘base set’ issue of the link graph. SALSA is based on the intuition that such an ‘authoritative’ page must be visible for thousands of pages of data set. Through a random walk of the graph, the algorithm indexes some authorities with relatively high probability. The theory of random walks combined the notion of hubs and authorities, which leads us to analyse two different Markov chains: the chain of hubs visited and the chain of authorities visited, which gives us, for each page, two distinct weights—the hubs and that of the authorities. To generate transition states of each of these chains, two edges of the graph must be traversed: the first forward (following an outgoing link) and the second backward (by following a link returning) or vice versa. Authorities weights are defined as the distribution of the stationary chain exploring in first random rearward link and then a forward link, while the weight of the hubs are defined as the distribution of the stationary chain exploring a first random forward link and then a backward link. More precisely, starting from a collection of hyperlinked documents, we can build the directed graph G. Let us consider Back(i) = { k: k → i}, the set of nodes that point to i , for example nodes that can be reached from i by following a link back, and let Forw (i) = { k: i → k }, the set of all nodes that can be reached starting from i by following a link to the before . | Back (i) | is the number of nodes that point to i, as | Forw (i) | is the number of nodes to which i points. We can now define two stochastic matrices, which contain the transition probabilities for Markov chains, respectively, for hubs and authorities.

The matrix for the hubs, H:

$$h_{i,j} = \sum_{k:k \in Forw(i) \cap Forw(j)} \frac{1}{|Forw(i)|} \frac{1}{|Back(k)|}$$

The matrix for the authorities, H:

$$a_{i,j} = \sum_{k:k \in Back(i) \cap Back(j)} \frac{1}{|Back(i)|} \frac{1}{|Forw(k)|}$$

Element ai, j > 0 means that at least one node k points to the two nodes i and j. The node j is reachable from the node i in two steps: the first going up the link i → k and the second by following the link k → j.

---

SALSA

---

**1** Initialize all weights to 1

**2** Repeat until the weights converge:

**3** for every hub $i \in H$

$$h_{i,j} = \sum_{k:k \in Forw(i) \cap Forw(j)} \frac{1}{|Forw(i)|} \frac{1}{|Back(k)|}$$

**4** for every *authority* $i \in A$

$$a_{i,j} = \sum_{k:k \in Back(i) \cap Back(j)} \frac{1}{|Back(i)|} \frac{1}{|Forw(k)|}$$

**5** Normalize

---

### 3.6 The PHITS algorithm

Other approaches for determining hubs and authorities were also tested. Cohn and Chang proposed a statistical algorithm to determine these two categories [17]. The model that the authors have constructed attempts to explain two types of variables, the quotes c of a document d, based on a small number of common variables z which are called aspects or factors. These common variables can be considered as subjects or community pages. The model can then be described statistically: a document d ∊ D is generated with a probability P (d), factor, or subject z∊ Z corresponding to d is selected in accordance with a probability P (z| d) , and since this factor, quotes c∊ C are generated based on the probability P ( c | z ). The probability of each pair (document, quote) (d, c) is then described by the following:

$$P(d,c) = P(d)P(c|d)$$

$$P(c|d) = \sum_{z} P(c|z)P(z|d)$$

Considering the matrix A representing pairs (document, quote), where the entry A [i, j] is non-zero if the document i has a link to the document j, the probability matrix A citation is as follows:

$$L(A) = \prod_{(d,c)\in A} P(d,c)$$

The problem then is to find the values of P (d), P (z | d) and P (c | z) that maximize the likelihood function L (A) of the observed data. To solve this new problem, the authors propose to use the EM algorithm of Dempster[18]. This fully probabilistic model has the advantage of providing more information than the model used by the HITS algorithm. An analogy can be made, however, considering the authorities on a given subject as the conditional probability P (c | z) which indicates how a document c is quoted from a community z. But other information can be extracted from the model such as the probability P (z | c), which allows us to know the community to which a given document c, or the discovery of documents that features a community in determining the product P (z | c) • P (c | z). Nevertheless, this algorithm imposes to know in advance the number of factors z to take into account. In addition, it is possible that the EM algorithm can get stuck in a local maximum and compromising convergence to the global maximum corresponding to the solution of the problem.

### 3.7The hub-averaging (HUBAVG) algorithm

Another set of algorithms that try to eliminate some of the drawbacks of HITS is also proposed by [11]. Thus, the ‘Hub-Averaging-Kleinberg’ algorithm is a combination of HITS and SALSA in which it tries to reduce the TKC effect (*Tightly-Knit Community Effect*). The calculation of scores of authority is the same as HITS, but the hub score is the average of scores of authority. The principle of the algorithm is that a page is a good hub (authority) if it links to (referenced by) good authorities (hubs) and scores hub (authority) are calculated by considering only the scores of authority (hub) that are greater than or equal to the average score.

$$a_i = \sum_{j \in B(i)} h_j \quad \text{and} \quad h_i = \frac{1}{|F(i)|} \sum_{j \in F(i)} a_j .$$

## 4 The Methodology

In this section, we present theprocess of data collection and the rank of the nodes using link analysis algorithms to detect communities and their members. Figure 1 represents the architecture of our work. It consists of four components: (i) the extraction of profiles with their links, (ii) the ranking of profiles according to their importance, (iii) the verification of profiles with higher rank if they have outgoing links, a page of community must not have outgoing links towards other nodes because the administrator can not send friendship requests from the community page, and (iv) the detection of communities and their members.

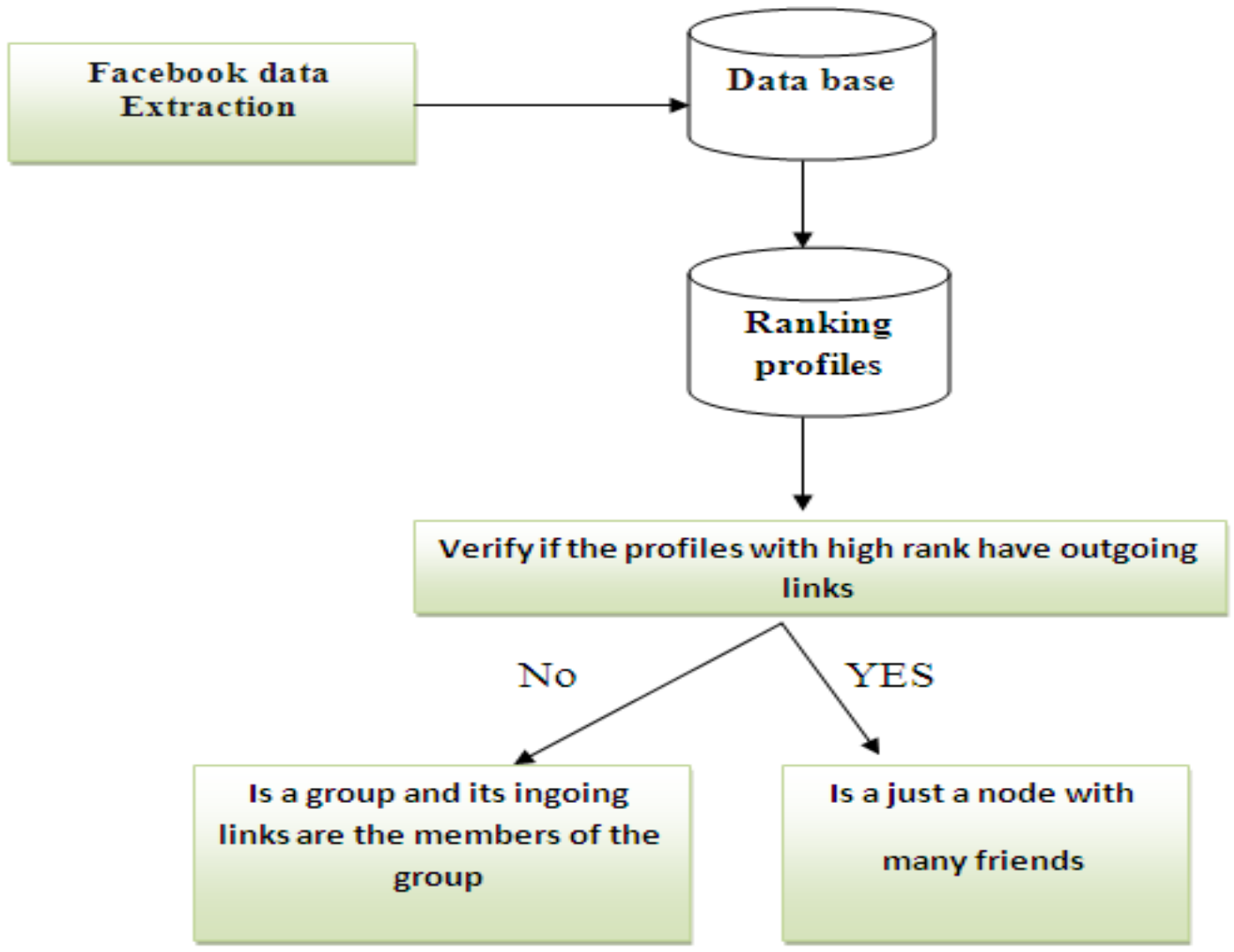

**Fig1** Approaches for detecting groups

### 4.1 Data collection

The set of data which have been used are obtained using Facebook. It later launched its API in May 2007 to attract those who are interested in the development of web applications. This API is available in numerous programming languages and it provides developers access to a vast quantity of information on the profiles of users. During this work, we have reached 1200 profiles on Facebook randomly with their links and friends to define the links.

All data are presented in the form of objects and connections

- **Objects**: persons, events, pictures, pages, groups, messages, .etc.
- **Connections**: friendship, shared content, like, etc.

Facebook will allow us to access these objects and then use the connections and get access to other objects. For this, the query will be constructed using an URL and answers will be returned in XML format. In simple terms, Facebook has only one entry point:

http://graph.facebook.com

The course of the SocialGraphis then just as simple. This will be donebyan object identifier and a connection definition:

http://graph.facebook.com/identifiant/connexion

The identifier is used to uniquely define an object in Facebook. Facebook member scan choose a string to create an alias identifiable by humans. For example, ausersimply chooseshis name "Jean-M. cornier"

http://graph.facebook.com/jean-M.cornier

We can see that this query returns:

{"name": " jean-M.cornier ",
"picture": "http://profile.ak.fbcdn.net/hprofile-ak
snc4/50270_68310606562_2720435_s.jpg",
"link": "http://www.facebook.com/ jean-M.cornier
","category": "Public figure",
"likes": 3711466,
"website": "www.facebook.com",
"username": " jean-M.cornier ",
"personal_interests": "openness, football, natation,
swimming, information flow, minimalism\n\n\n"}

Connection: we can then leave this identifier and the object it refers to access other objects through connections. Here are some examples:

To access the list of friends of a person:

http://graph.facebook.com/jean-M.cornier/friends

To access the list of videos posted by a person

http://graph.facebook.com/ jean-M.cornier /videos

To access the list of photos posted by a person

http://graph.facebook.com/jean-M.cornier/pictures

### 4.2 Content extraction

The links of each node are extracted by the Facebook API according to the characteristics of the social network structures; we have composed the content as a matrix of links. The matrix is shown below.

Matrice =

| | Nœud(1) | nœud(2) | nœud(3) | nœud(4)......nœud(n) |
|---|---|---|---|---|
| Nœud(1) | 0 | 0 | 1 | 0 |
| Nœud(2) | 1 | 0 | 0 | 1 |
| Nœud(3) | 0 | 1 | 0 | 0 |
| Nœud(4) | 0 | 0 | 1 | 0 |

The matrix will be used for ranking. The links will be treated as parameters of link analysis algorithms PAGERANK, HITS, PHITS and SALSA. The experiment on performance approach measure will be discussed in the next section.

## 5 Experimental Results

In order to evaluate the prediction performance, we designed a series of experiments with the four algorithms. Figure 2 shows the experimental conception of this work. The results of this evaluation are shown in Table1.

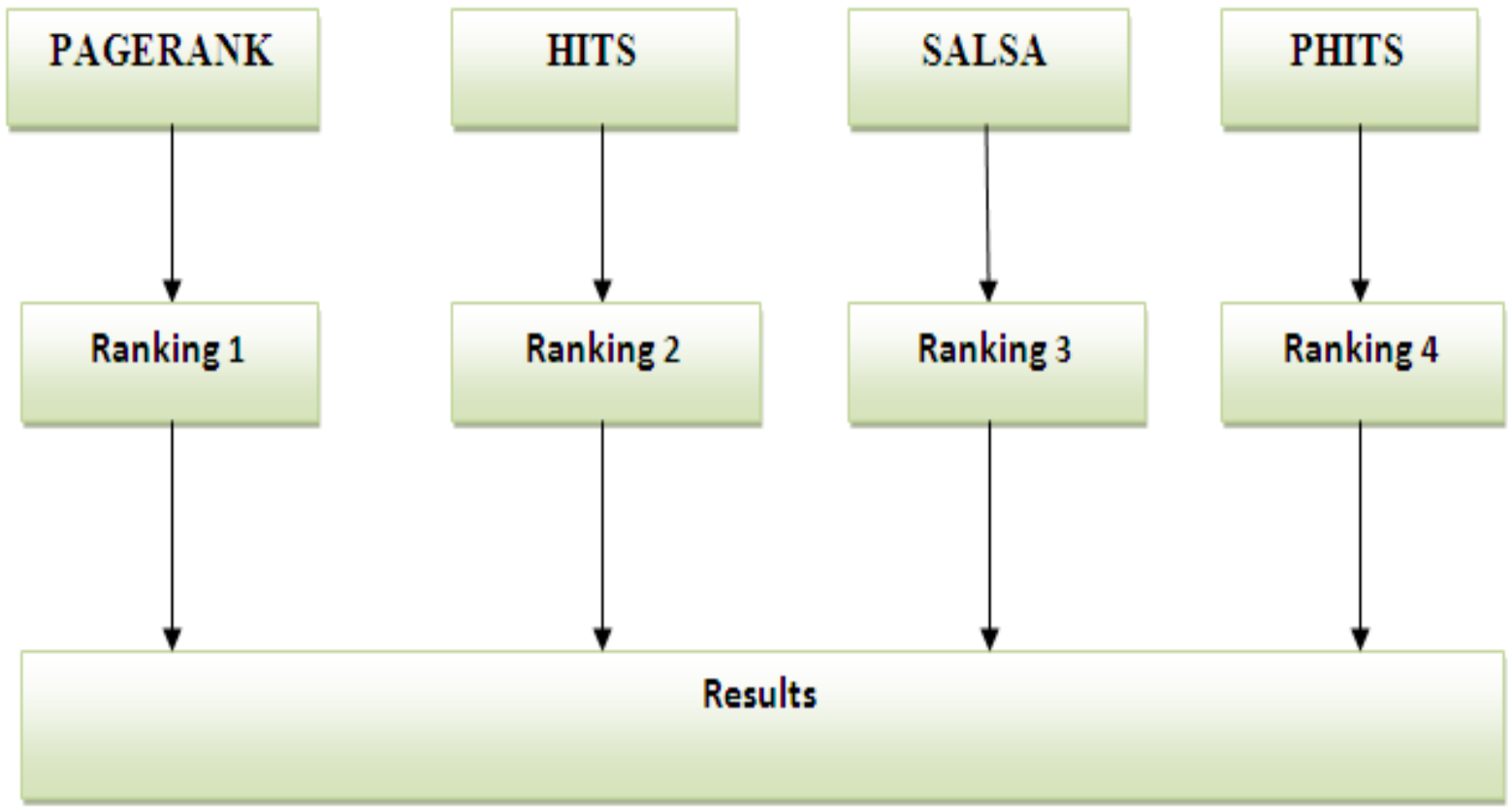

**Fig. 2 Groups detection**

A social network is seen as a dynamic structure presented with nodes and links. The nodes are generally designed by individuals or organizations and they are connected by social interactions. The visualization method that we have used is JAVA3D which proposes a general view with the possibility to develop a nod so as to reach the details of its profile. Figure 3 shows the data visualization.

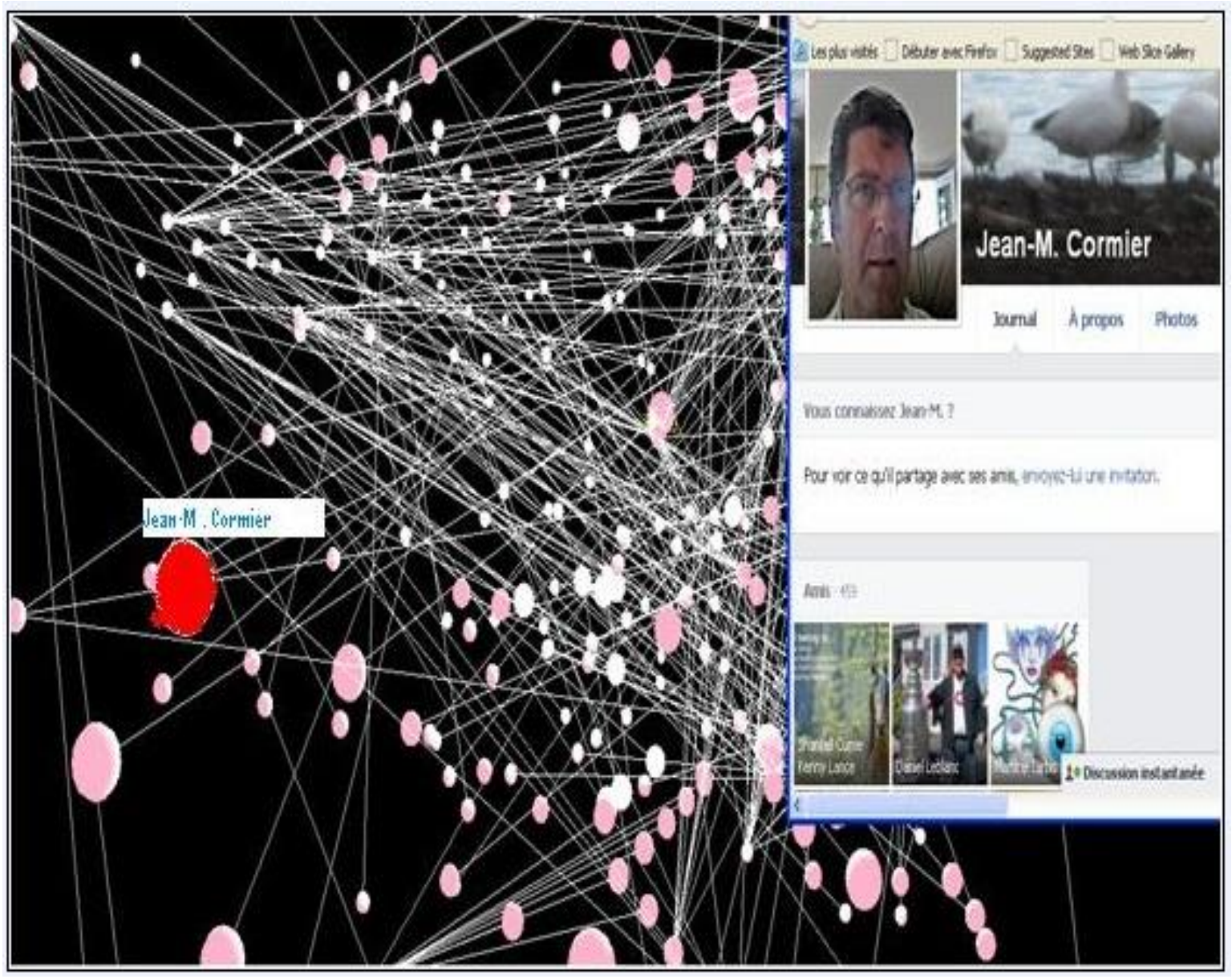

**Fig. 3 Data visualization**

Concerning the algorithms HITS, PHITS and SALSA, we take just the values of authorities as values of important nodes and we consider the role of the hubs that indicates the authorities. The results in Table 1 are for the nodes which have higher rank.

**Table 1:** Performance evaluation summarization table of groups detection with PageRank

| **PageRank** | | **HITS** | | | **SALSA** | | | **PHITS** | | |
|---|---|---|---|---|---|---|---|---|---|---|
| ***RI*** | ***Groups and their*** | ***hubs*** | ***Autho*** | ***Groups and their*** | ***Hubs*** | ***Autho*** | ***Groups and their*** | ***Hubs*** | ***Autho*** | ***Groups and their*** |

| | ***members*** | | ***rities*** | ***members*** | | ***rities*** | ***members*** | | ***rities*** | ***members*** |
|---|---|---|---|---|---|---|---|---|---|---|
| 0.801<br>0.733<br>0.721<br>0.701<br>0.699<br>0.655<br>0.652<br>0.652<br>0.633<br>0.601<br>0.573<br>0.571<br>0.569<br>0.547<br>0.5<br>0.487<br>0.466 | Total number of groups (12)<br>G1 (65) member<br>G2 (122) member<br>G3 (61) member<br>G4 (65) member<br>G5 (99) member<br>G6 (152) member<br>G7 (200) member<br>G8 (50) member<br>G9 (42) member<br>G10 (36) member<br>G11 (36) member<br>G12 (20) member | 0.901<br>0.899<br>0.897<br>0.892<br>0.891<br>0.884<br>0.881<br>0.866<br>0.862<br>0.834<br>0.822<br>0.805<br>0.799<br>0.782<br>0.736<br>0.723<br>0.712 | 0.821<br>0.826<br>0.831<br>0.839<br>0.843<br>0.862<br>0.866<br>0.869<br>0.875<br>0.879<br>0.881<br>0.888<br>0.889<br>0.9<br>0.9<br>0.9<br>0.901 | Total number of groups (10)<br>G1 (70) member<br>G2 (123) member<br>G3 (42) member<br>G4 (77) member<br>G5 (77) member<br>G6 (159) member<br>G7 (201) member<br>G8 (60) member<br>G9 (67) member<br>G10 (72) member | 0.921<br>0.901<br>0.889<br>0.881<br>0.805<br>0.795<br>0.782<br>0.781<br>0.766<br>0.762<br>0.747<br>0.741<br>0.739<br>0.711<br>0.709<br>0.701<br>0.7 | 0.912<br>0.9<br>0.899<br>0.898<br>0.895<br>0.891<br>0.888<br>0.884<br>0.881<br>0.881<br>0.881<br>0.878<br>0.877<br>0.874<br>0.871<br>0.868<br>0.859 | Total number of groups (09)<br>G1 (71) member<br>G2 (134) member<br>G3 (46) member<br>G4 (84) member<br>G5 (100) member<br>G6 (171) member<br>G7 (206) member<br>G8 (66) member<br>G9 (80) member | 0.691<br>0.688<br>0.683<br>0.676<br>0.655<br>0.642<br>0.6<br>0.599<br>0.599<br>0.589<br>0.587<br>0.581<br>0.578<br>0.571<br>0.564<br>0.561<br>0.5 | 0.688<br>0.681<br>0.641<br>0.615<br>0.602<br>0.601<br>0.545<br>0.532<br>0.511<br>0.487<br>0.477<br>0.468<br>0.466<br>0.459<br>0.455<br>0.449<br>0.405 | Total number of groups (07)<br>G1 (102) member<br>G2 (151) member<br>G3 (50) member<br>G4 (98) member<br>G5 (161) member<br>G6 (180) member<br>G7 (206) member |

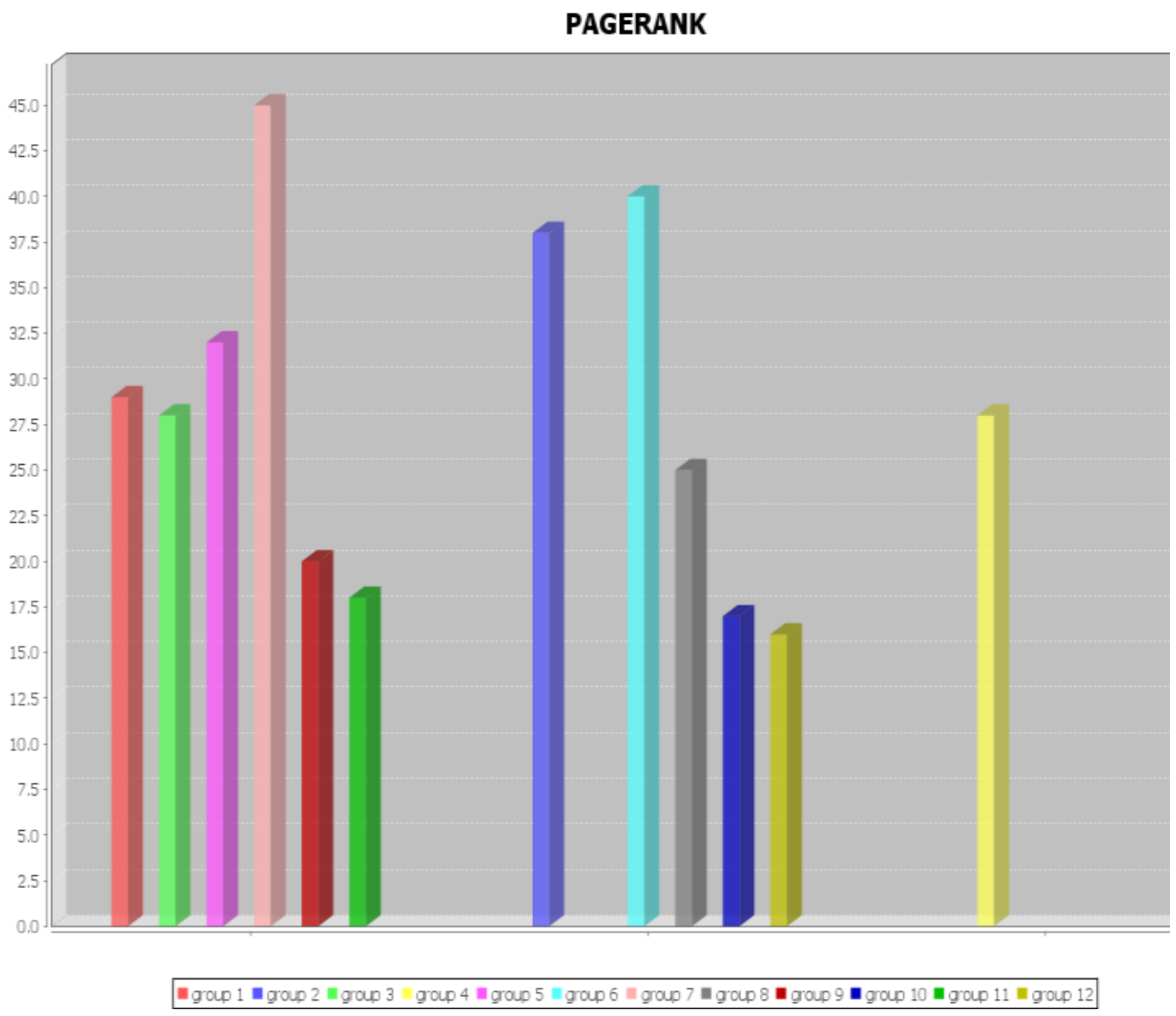

**Fig4** Groups and their members detected by PageRank

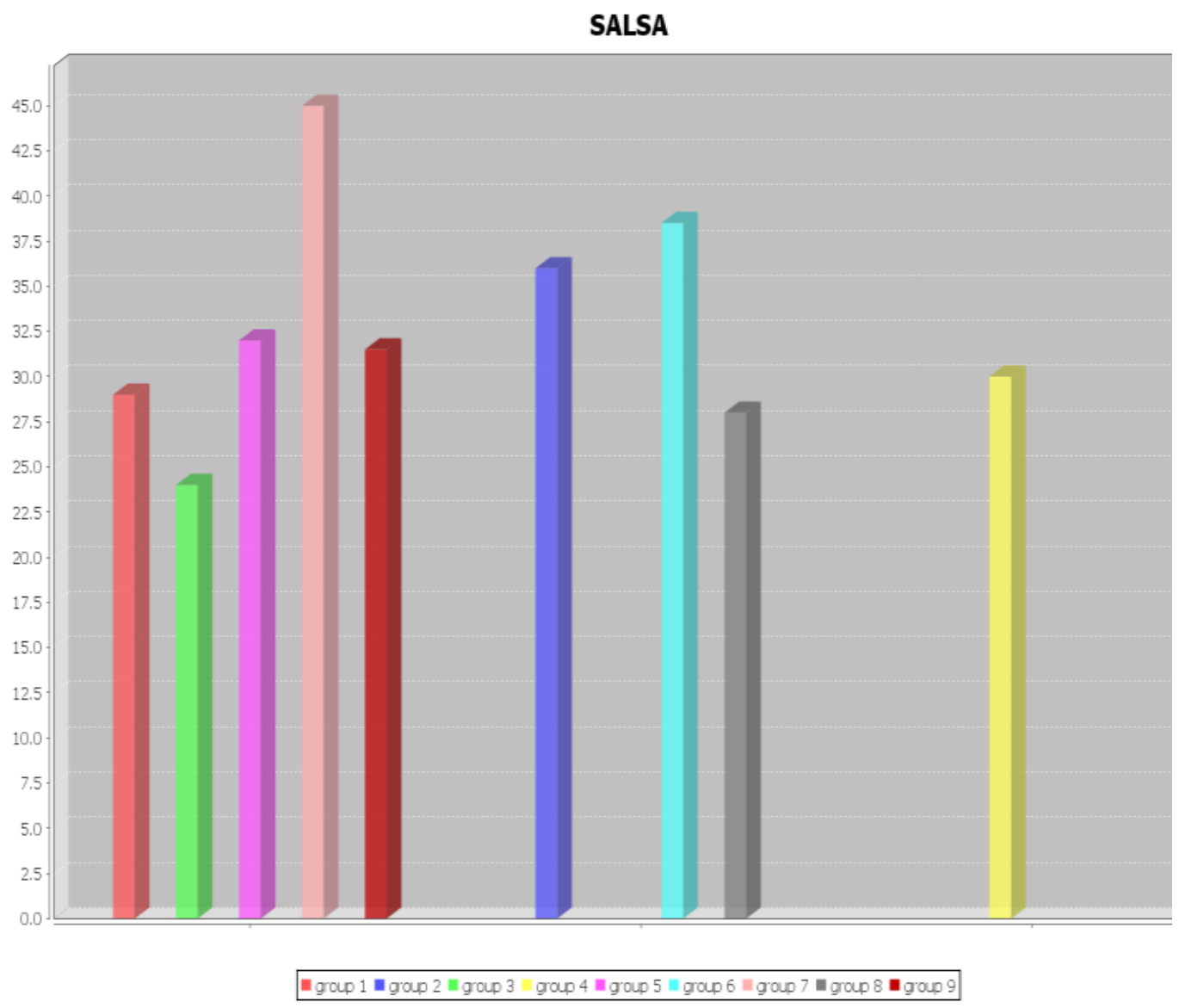

**Fig6** Groups and their members detected by SALSA

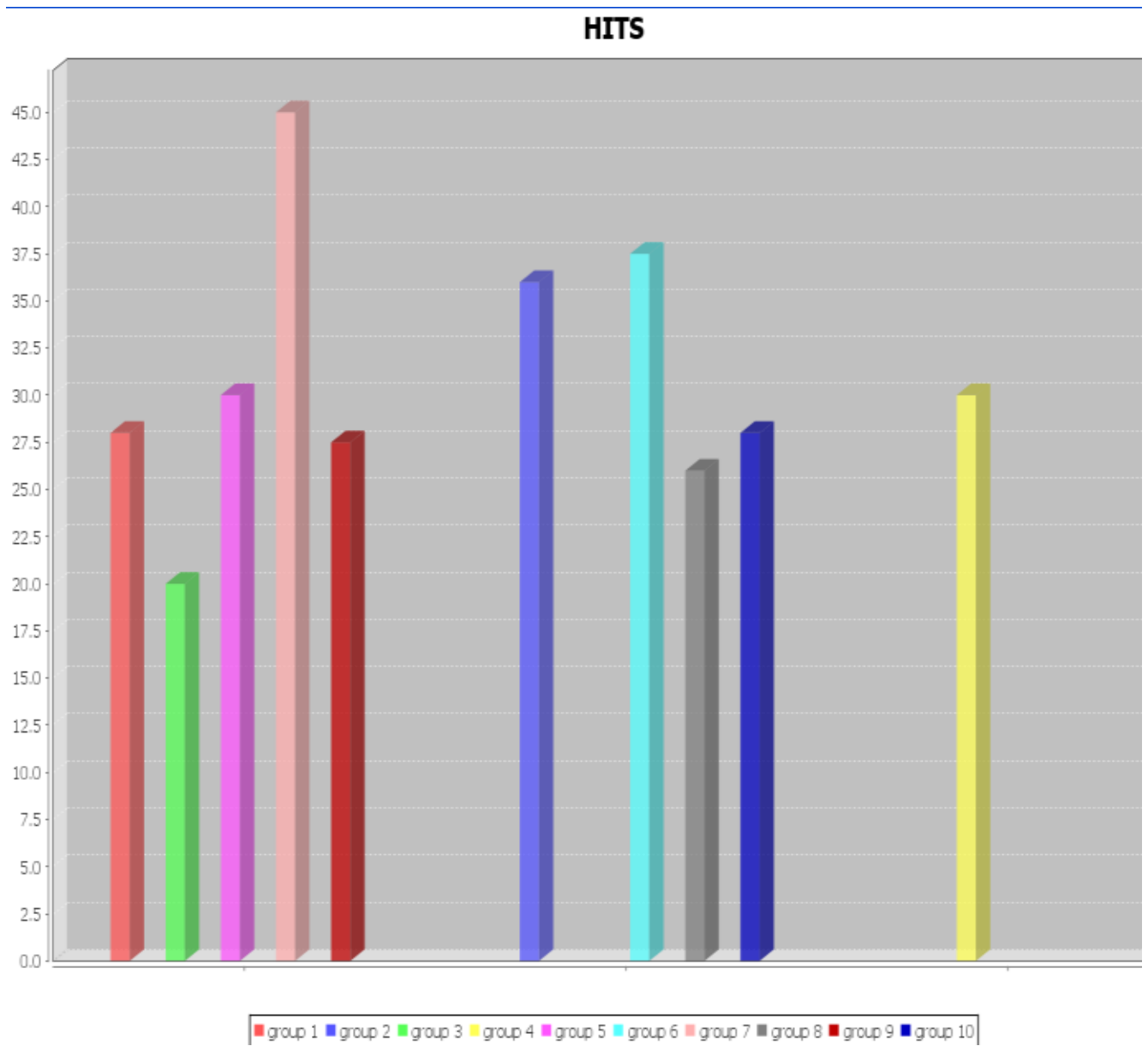

**Fig5** Groups and their members detectedby Hits

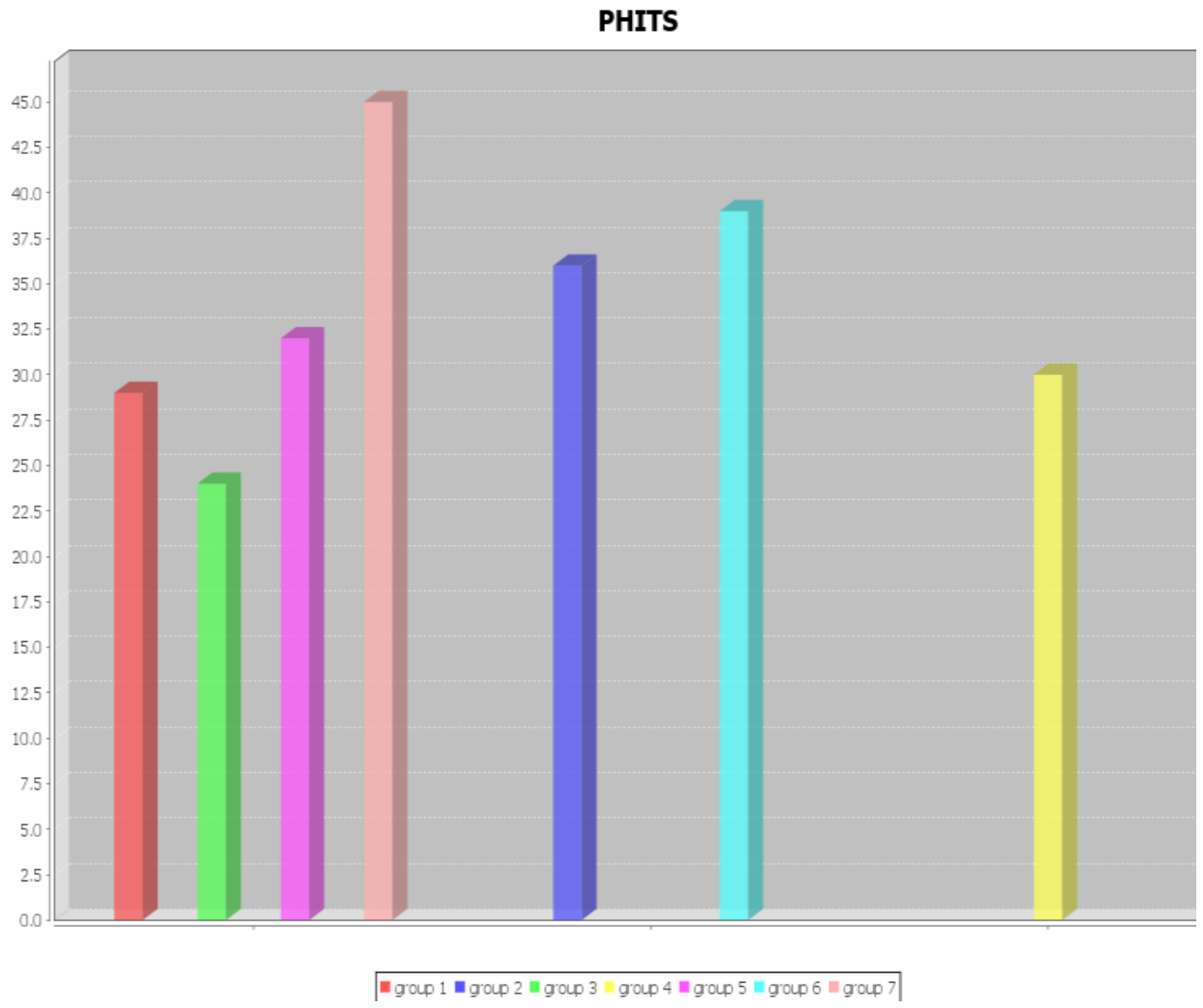

**Fig7** Groups and their members detected by PHITS

Concerning PAGERANK algorithm, the existence of communities which are heavy with hubs and authorities is meaningless for the algorithm because it is not based on a mutual reinforcement to calculate the weight of authority. We remark that PAGERANK attributes an important weight to the isolated node with the maximum degree. Generally, we observe that PAGERANK favours the isolated nodes with high grade, and the hubs which point towards the isolated node transfer all their weights directly to this node augmenting its weight. PAGERANK detects the overlapping between the communities; the ability of the algorithm in detecting overlapping between communities is due to the recursive leaps which are performed by PAGERANK. The recursive leaps are probably responsible which enable the algorithm to be the best among the other algorithms that we used in this work.

We also observe that the results of PHITS are low and have shown the weakness of this algorithm. The results also show that HITS and SALSA have nearly the same performance with a little advancement of HITS.

## 6Conclusion

We conducted an experimental analysis ranking with link analysis by evaluating the ability of each algorithm to classify the user profiles on Facebook, and its ability to affect the existence of communities in a graph. We observed that PAGERANK is the most efficient.

We plan to apply these algorithms to detect communities in other social networks such as Twitter, Plurk, and Blogger, etc., and subsequently to develop approaches that detect the behaviour of members of different groups.